# Interfacial Dzyaloshinskii-Moriya interaction in Ta\Co$_{20}$Fe$_{60}$B$_{20}$\MgO nanowires

R. Lo Conte,[1,2] E. Martinez,[3] A. Hrabec,[4] A. Lamperti,[5] T. Schulz,[1] L. Nasi,[6] L. Lazzarini,[6] F. Maccherozzi,[7] S. S. Dhesi,[7] B. Ocker,[8] C. H. Marrows,[4] T. A. Moore,[4] and M. Kläui[1,2]

[1]Johannes Gutenberg Universität-Mainz, Institut für Physik, Staudinger Weg 7, 55128 Mainz, Germany
[2]Graduate School of Excellence Materials Science in Mainz (MAINZ), Staudinger Weg 9, 55128 Mainz, Germany
[3]Dpto. Fisica Aplicada, Universidad de Salamanca, plaza de los Caidos s/n E-38008, Salamanca, Spain
[4]School of Physics and Astronomy, E. C. Stoner Laboratory, University of Leeds, Leeds LS2 9JT, U.K.
[5]Laboratorio MDM, IMM-CNR, via C. Olivetti 2, 20864 Agrate Brianza (MB), Italy
[6]IMEM-CNR, Parco Area delle Scienze 37/A, 43124 Parma (PR), Italy
[7]Diamond Light Source, Chilton, Didcot, Oxfordshire, OX11 0DE, U.K.
[8]Singulus Technologies AG, 63796 Kahl am Main, Germany

We report current-induced domain wall motion (CIDWM) in Ta\Co$_{20}$Fe$_{60}$B$_{20}$\MgO nanowires. Domain walls are observed to move *against* the electron flow when no magnetic field is applied, while a field along the nanowires strongly affects the domain wall motion direction and velocity. A symmetric effect is observed for up-down and down-up domain walls. This indicates the presence of right-handed domain walls, due to a Dzyaloshinskii-Moriya interaction (DMI) with a DMI coefficient D=+0.06 mJ/m$^2$. The positive DMI coefficient is interpreted to be a consequence of boron diffusion into the tantalum buffer layer during annealing. In a Pt\Co$_{68}$Fe$_{22}$B$_{10}$\MgO nanowire CIDWM *along* the electron flow was observed, corroborating this interpretation. The experimental results are compared to 1D-model simulations including the effects of pinning. This advanced modelling allows us to reproduce the experiment outcomes and reliably extract a spin-Hall angle θ$_{SH}$=-0.11 for Ta in the nanowires, showing the importance of an analysis that goes beyond the currently used model for perfect nanowires.

## I. INTRODUCTION

The increasing demand for data storage devices able to store information with higher density has led to an enormous effort investigating materials systems useful for such a purpose. In information and communication technology magnetic materials are used extensively [1]. Nowadays, scientific interest is moving from single magnetic materials-based systems [2] to more complicated heterostructures [3]. The latter are materials systems characterized by perpendicular magnetic anisotropy (PMA) and structural inversion asymmetry. The PMA results in domain wall (DW) widths of a few nm [4], which offer the opportunity of a high data-storage density. Examples of such materials stacks are Pt\Co\AlO$_x$ [5-8], Pt\[Co\Ni]$_x$\Co\TaN [3], Ta\CoFe\MgO [9] and Ta\CoFeB\MgO [10-13], which have a magnetization pointing out of the plane and no inversion symmetry in the vertical direction. Very effective current-induced domain wall motion (CIDWM) [3,5,7] and magnetization switching [14-17] have been observed in nanostructures made of such materials. After the first experimental observations, the Rashba effect [6,18] and the spin-Hall effect (SHE) [14,19,20] were considered to be the possible leading effects for the magnetization dynamics in such systems. More recent results support the interpretation that the SHE is likely to be the main cause [3,9]. According to the spin-orbit torque (SOT)-model, the symmetry of the resulting torque is defined by the SHE generated in the heavy metal underlayer and the interfacial Dzyaloshinskii-Moriya interaction (DMI) between the heavy metal and the magnetic layer [3,9,21,22]. In this scenario the DMI results in fixing the chirality of the domain walls [22].

Much attention has been dedicated to systems such as Ta\CoFeB\MgO, due to the fact that this materials stack is already used for the fabrication of spintronic devices [23] whose functionality is based on the spin-transfer torque (STT) [24]. Now that the SOT has been discovered, the new challenge is to understand what exactly governs the torques and the DMI strength and sign. Some steps forward in the understanding of the symmetry of the torque have been taken, measuring the angular dependence of the generated effective fields [12,13]. However, in those experiments only the mono-domain state of a magnetic nanostructure was probed. Instead, it is in the dynamics of domain walls that the DMI starts to play an important role, so that in order to learn more about such an interaction it is necessary to carry out DWM experiments.

Here we report CIDWM in out-of-plane magnetized Ta\Co$_{20}$Fe$_{60}$B$_{20}$\MgO nanowires. The DW velocity is measured in the presence of a variable external magnetic field applied along the wire axis. A strong effect of this longitudinal field on the DW motion is observed, allowing us to measure the DMI strength D for the hetero-structure under investigation. Diffusion and consequent segregation of boron at the Ta\CoFeB interface are found and play an important role in governing the interfacial DMI in our system. Observation of CIDWM along the electron flow in Pt\Co$_{68}$Fe$_{22}$B$_{10}$\MgO nanowires supports this interpretation. Comparing experiments to 1D-model simulations, we are able to interpret the role of the pinning in the DW dynamics and we also extract the value of the spin-Hall angle of Ta in the nanowires with high confidence.

## II. SAMPLES AND EXPERIMENTAL SET-UP

Our sample consists of Ta(5nm)\Co$_{20}$Fe$_{60}$B$_{20}$(1nm)\MgO(2nm)\Ta(5nm) deposited on a thermally oxidized Si-wafer. The entire materials stack is deposited by sputtering (using a Singulus TIMARIS/ROTARIS tool), and then annealed at 300 °C for 2 hours in vacuum so that a large PMA is obtained. For a measured in-plane magnetization saturation field $\mu_0 H_{sat}$=400 mT and a saturation magnetization $M_s$=1.1x10$^6$ A/m (measured by SQUID), we obtain an effective anisotropy $K_{eff}=\mu_0 H_{sat} M_s/2$=2.2x10$^5$ J/m$^3$. By electron-beam lithography and argon-ion milling the sample is then patterned into an array of 20 nanowires in a parallel geometry (see Fig. 1(a)). The dimensions of each wire are 1 μm x 8 μm. At the ends of the wires there are magnetic pads, directly connected to two gold contact pads made in a second patterning step by a lift-off technique. One of the two gold pads consists of an Oersted-line, used for the nucleation of reversed magnetic domains in pre-saturated wires, by the injection of 20 ns-long current pulses. Furthermore, we also use a sample made of Pt(4.5)\Co$_{68}$Fe$_{22}$B$_{10}$(0.6)\MgO(1.5) (thicknesses in nm), and more details about this sample are reported in section IV.

As shown in Fig. 1(a), a pulse generator is used for injecting current through either the Oersted-line or the magnetic wires. An oscilloscope is used for measuring the pulse waveform, across its 50 Ω-internal resistance ($R_o$). The total current flowing through the system is obtained by the measured voltage $V_o$ across $R_o$. Taking into account the oxidation of the top 2 nm of the Ta capping layer (see Fig. 5(a)), we estimate a current density of $1.1 \times 10^{11}$ A/m² flowing through the nanowires when 1 V drops across $R_0$ (corresponding to a total current of 20 mA). The conventional current density $j_a$ is assumed to be positive when it flows in the +x-direction (see Fig. 1), so that the electron current density $j_e<0$ is in the +x-direction. The magnetization state of the wires is imaged by polar Kerr microscopy in differential mode. A magnetic coil is used for the generation of an external in-plane magnetic field. The experiments are carried out at T=300 K. We first saturate the magnetic wires in the up (+z)- or down (-z)-magnetization state, then we nucleate reversed domains either by the Oersted-line or injecting a current pulse through the wires. By the injection of an Oersted-pulse we generate only one type of domain wall (DW), either up-down (↑↓)- or down-up (↓↑)-DW. On the other hand, we generate both types of DWs by injecting current through the nanostructures, due to current-induced magnetization switching [17]. Fig. 1(b) and 1(c) show controlled domain nucleation by current through the Oersted-line and DW displacement due to the injection of a burst of negative current pulses ($j_a<0$) in the wires, respectively.

### III. CURRENT-INDUCED DOMAIN WALL MOTION

Fig. 2 reports the average velocity of the DW as a function of the current density flowing through the magnetic wires. For each current density the measurement is repeated three times, yielding a total of 30 DW displacements (10 nanowires are imaged at the same time). This allows us to obtain sufficient statistics for the DW motion. Bursts of several (*n*) current pulses with a time duration *Δt*=10, 15, 20 and 25 ns are used for the CIDWM (see Fig. 2(a)). The time between two consecutive pulses is 100 μs and the number of pulses in a burst ranges from *n*=20 to *n*=400. The velocity of the DW is calculated as the ratio between the displacement of the domain wall due to the injected pulse burst and the total pulsing time *T=n\*Δt*. With the measurements for the different pulse lengths it is possible to rule out the effect of the rise and fall-time (5 ns each in our experimental set-up) on the measured DW velocity. Indeed, as shown in Fig. 2(a), higher velocities are measured for longer pulse lengths at a fixed current density. This is because the rise/fall time takes up a smaller proportion of the overall pulse length $Δt$, enabling the torque on the DW to be larger for a greater fraction of the chosen time. This has to be taken into account when comparing results for different $Δt$. In Fig. 2(b) the resulting average DW velocity free of the influence of the rise- and fall-time is shown. One of the key information in Fig. 2 is the direction of the DW velocity: the DWs move against the electron flow. This is a clear indication of the fact that in our system the DWs are not moved by conventional STT [24], which would move them in the electron flow direction. Instead, the observed DWM is in agreement with the SOT-model [3,9,22]. A similar interpretation was given for Pt\CoFe\MgO and Ta\CoFe\MgO systems [9], where the authors claimed that the DWM is due to the SHE-effective field $\boldsymbol{H}_{SHE} = \hbar\theta_{SH}|j_e|/(2\mu_0|e|M_s L_z)(\hat{\boldsymbol{m}} \times (\hat{\boldsymbol{z}} \times \hat{\boldsymbol{j}_e}))$, where $θ_{SH}$ is the spin-Hall angle (SHA), $j_e$ is the electron-current density, $M_s$ is the saturation magnetization of the ferromagnetic material and $L_z$ is the thickness of the ferromagnetic layer. However, in our experiments DWM in the opposite direction compared to the case of Ta\CoFe\MgO in [9] is observed. The main difference between the materials system in [9] and the stack investigated here is the presence of boron (B) in the ferromagnetic layer. The origin of the observed DWM in the opposite direction will be discussed below.

## IV. CHIRAL DOMAIN WALLS

In the SOT-model, the driving force for the DW dynamics is the pure spin-current induced by the SHE generated in the heavy metal during the pulse injection. Furthermore, the DMI at the interface between the heavy metal and the ferromagnetic layer is responsible for the initial magnetic configuration of the DW [22,25], fixing its final direction of motion. As a consequence, the direction of the DWM depends on both the sign of the SHA and the sign of the DMI, where the latter fixes the chirality (left- or right-handed) of the Néel-component of the DW. Regarding the SHA of Ta, it is known to have a negative sign, as reported in the literature for different materials stacks [9,26] as well as for the very same materials system as used here [17]. Meanwhile, little has been reported on the DMI in Ta\CoFeB\MgO systems so far [27].

The effect of magnetic fields on CIDWM is now investigated. More precisely, the DW velocity is measured as a function of an applied magnetic field along the wire axis (x-direction) for fixed current densities (see Fig. 3). First of all, both types of DW (↑↓ and ↓↑) are nucleated in the pre-saturated nanostructures by current-induced magnetization switching (see Fig. 3(a)). Typical nucleation pulses used in the experiment have a current density $j_a$~$10^{12}$ A/m$^2$ and a duration $\Delta t$=20 ns. Once the DWs are generated, they are displaced by injecting a burst ($n$=1-20) of 20 ns-long current pulses with lower current densities (2.8-3.6x10$^{11}$ A/m$^2$), as shown in Fig. 3(b). In order to calculate the DW velocity, the full width at half maximum of the current pulse is used as the time duration of a single pulse. For each current density-field amplitude combination the measurement is repeated five times. The DW velocity as a function of the longitudinal field $\mu_\circ H_x$ is shown in Fig. 3(c) and 3(d) (symbols), respectively for the ↓↑- and ↑↓-DWs. The graphs show that the DW velocity is strongly influenced by the presence of the longitudinal field.

While at zero field the velocity of both types of DWs is the same, in the presence of the field the two DWs move at different velocities. We observe a symmetric behavior of the DW velocity as a function of $\mu_\circ H_x$ for the two types of DW, as shown in Fig. 3(e) (solid symbols for the ↑↓-domain wall, empty symbols for the ↓↑-domain wall). Similar observations have been reported for magnetic structures made of different materials stacks [3,9]. The field at which the SOT is minimized, resulting in a stationary DW, is the so-called DMI effective field $\mu_o H_{DMI}$=$D/(M_S \Delta)$ [3,28], where $D$ is the DMI coefficient, and $\Delta$ is the DW width. Fig. 3(c), (d) and (e) show that there is a range of in-plane longitudinal fields $\mu_\circ H_x$ where the DW remains stationary (with zero or very small DW velocity compared to the velocities measured for larger longitudinal fields). This zero motion field range is not reproduced by the standard SOT-DWM model and was not discussed in some other experiments [3,9]. As shown later, in order to properly analyze the experimental data a more accurate model is needed, where this "pinning" effect is taken into account.

Since the reversal of the direction of the DW motion occurs for the low-velocity field range, a more detailed analysis of this behavior follows. The DMI-field is extracted by linearly fitting the experimental data in Fig. 3(e), for both types of DW and for both positive and negative current.

Considering only the high velocity experimental data, the crossing of the two best fitting lines for the ↑↓-DW data occurs at a longitudinal field value $\mu_oH_x^{↑↓}$=-8.5±1.8 mT. While, for the ↓↑-DW the crossing occurs at $\mu_oH_x^{↓↑}$=+7.0±1.5 mT. Assuming the amplitude od the DMI field to be the average of the two fields (in absolute values) we obtain $|\mu_oH_{DMI}|$=7.8±1.2 mT. All the errors correspond to one standard deviation. Since the DW width is Δ=7 nm (Δ=$(A/K_{eff})^{1/2}$, where we use A=$10^{-11}$ J/m [28]), $\mu_oH_x^{↑↓}$<0 and $\mu_oH_x^{↓↑}$>0, and knowing that Ta-$\theta_{SH}$ has a negative sign [9,17,26] we obtain a DMI constant D=+0.06±0.01 mJ/$m^2$. Such a value is close to the one measured for the Ta\CoFe interface [28], but of opposite sign. This indicates the presence of right-handed DWs in our nanowires, while left-handed DWs were reported for Ta\CoFe\MgO nanowires [28].

To determine the origin of this difference, we replace Ta by Pt and we investigate CIDWM in a Pt(4.5)\Co$_{68}$Fe$_{22}$B$_{10}$(0.6)\MgO(1.5) nanowire (thicknesses in nm). In this experiment we have a single 500 nm-wide magnetic wire with a magnetic pad on one side, connected to two gold pads at its ends (see Fig. 4). The technique used for imaging the magnetic domains is x-ray magnetic circular dichroism combined with photoemission electron microscopy (XMCD-PEEM). This technique is used due to its higher spatial resolution compared to Kerr microscopy, which can help to image smaller structures and the magnetic texture with a higher resolution. As shown in Fig. 4, DWs are observed to move along the electron flow when a burst of current pulses ($j_a$=5x$10^{11}$ A/$m^2$) is injected in the nanowire. According to the SOT-model, such a movement can only be explained by the presence of right-handed domain walls (D>0), since the spin-Hall angle of Pt is positive [9]. This means that D has an opposite sign with respect to the DMI coefficient reported for nanowires made of Pt\CoFe\MgO [9], corroborating the idea that the B contained in the system governs the sign of the DMI.

## V.  BORON DIFFUSION AND SEGREGATION

In the search of an explanation for the positive DMI coefficient observed here, it is important to note that the Ta\Co$_{20}$Fe$_{60}$B$_{20}$\MgO stack was annealed at 300 °C for 2 h in vacuum. This was done in order to obtain a strong PMA. Transmission electron microscopy (TEM) imaging and secondary ion mass spectrometry (SIMS) are used in order to investigate the structural properties of the materials stack. The TEM cross-section image in Fig. 5(a) confirms the nominal thickness of each layer composing the stack, and provides evidence for the presence of a top Ta-oxide layer due to the natural Ta oxidation in air. The interfaces are smooth and well defined; in particular the MgO\Ta interface retains sub-nanometer topographical roughness after the crystallization of the MgO and CoFeB layers upon annealing, in agreement with other reported results [29]. The TEM image also shows the crystallization of MgO and CoFeB. On the other hand, time-of-flight SIMS depth profiling in Fig. 5(b) on exactly the same materials stack clearly shows that B diffuses out of the CoFeB layer into the MgO and into the Ta-layer during the annealing process. Considering, in detail, the B profile in Fig. 5(b) for as-deposited and annealed samples, it is observed that the profile related to the annealed sample shows: *(i)* a reduced B intensity in the CoFeB layer; *(ii)* an increased B intensity in the region corresponding to the underneath Ta layer; *(iii)* a different modulated intensity close to the MgO\CoFeB interface. This is direct evidence for B diffusion from CoFeB to the adjacent layers, thus affecting in particular CoFeB\Ta interface where even B segregation is expected [30]. Since the DMI is expected to be a function of the structural and the atomic configuration at the interface [21,25], we can attribute the positive sign of D to the B diffusion and segregation in the Ta buffer layer, as this is the key difference to the stacks with CoFe instead of CoFeB.

## VI. DMI AND SHE EXTRACTION BY ADVANCED 1D-MODELLING

To quantify the DMI and the SHE we analyze the DW velocities in Ta\Co$_{20}$Fe$_{60}$B$_{20}$\MgO nanowires shown in Fig. 3(c)-(e) more in details. As stated above, there is a range of longitudinal magnetic fields for which the DW stops moving or it moves with a very low average velocity. However, when the longitudinal field reaches a certain value the domain wall velocity increases suddenly. Here an interpretation of such observations is offered, based on a 1D-model including DW pinning effects.

In the framework of the 1D-model (1DM), the DW dynamics is described in terms of the DW position $X$ and the DW angle $\Phi$ by the following equations [9,31]

$$(1 + \alpha^2) \frac{1}{\Delta} \frac{dX}{dt} = Q \Omega_A + \alpha \Omega_B \tag{1}$$

$$(1+\alpha^2)\frac{d\Phi}{dt} = -\alpha\Omega_A + Q\Omega_B \quad (2)$$

with

$$\Omega_A = -\frac{\gamma_0}{2}H_K\sin(2\Phi) + \gamma_0\frac{\pi}{2}H_x\sin(\Phi) - Q\gamma_0\frac{\pi}{2}H_{DMI}\sin(\Phi)$$

$$\Omega_B = Q\gamma_0\frac{\pi}{2}H + Q\gamma_0\frac{\pi}{2}H_{SHE}\cos(\Phi),$$

where $\Delta=(A/K_{eff})^{1/2}=7$ nm is the DW width with $K_{eff}=K_u-\mu_\circ M_s^2/2$, where $K_u=9.8\times10^5$ J/m$^3$ is the PMA anisotropy constant, $A=10^{-11}$ J/m is the exchange constant and $M_s=1.1\times10^6$ A/m the saturation magnetization. These inputs correspond to the experimentally deduced value for $K_{eff}=2.2\times10^5$ J/m$^3$. $H_K=N_xM_s$ is the shape anisotropy field with $N_x=L_zLog(2)/(\pi\Delta)$ being the magnetostatic factor [32]. $\alpha=0.013$ is the Gilbert damping [33]. $L_z=1$ nm is the thickness of the ferromagnetic layer and $L_y=1000$ nm its width. The factor $Q=+1$ and $Q=-1$ for the ↑↓ and ↓↑ configurations respectively. In the framework of the 1DM, the DMI generates an effective field along the x-axis with amplitude given by [22] $H_{DMI} = \frac{D}{\mu_0 M_s \Delta}$, where $D$ is the DMI parameter. $H_{SHE}$ is the effective spin-Hall field given by [34] $H_{SHE} = \frac{\hbar\theta_{SH}j_a}{2e\mu_0 M_s L_z}$, where $\theta_{SH}$ is the spin Hall angle, $e$ is the electron charge and $j_a$ is the current density ($\mathbf{j_a}=j_a\mathbf{u_x}$ and $\mathbf{j_e}=j_e\mathbf{u_x}$, with $j_a>0$ along the +x-direction and $j_e=-j_a$). $H_x$ is the applied longitudinal field along the x-axis, and $H=H_{pin}+H_{th}$ includes the pinning field $H_{pin}(X)$ and the thermal field $H_{th}$. The spatially-dependent pinning field accounts for local imperfections (such as edge or surface roughness or defects), and can be derived from an effective spatially-dependent pinning potential $V_{pin}(X)$, thus [35]

$$H_{pin}(X) = -\frac{1}{2\mu_0 M_s L_y L_z}\frac{\partial V_{pin}}{\partial X}.$$

A periodic potential was assumed to describe the experimental results

$$V_{pin}(X) = V_0\sin(\frac{\pi X}{p}),$$

where $V_0$ is the energy barrier of the pinning potential and $p$ is its spatial period. Finally, the thermal field $H_{th}(t)$ describes the effect of thermal fluctuations, and it is assumed to be a random Gaussian-distributed stochastic process with zero mean value ($<H_{th}(t)>=0$), uncorrelated in time ($<H_{th}(t)H_{th}(t')> = \frac{2\alpha K_B T}{\gamma_0 \mu_0 M_s \Delta L_y L_z} \delta(t-t')$), where $K_B$ is the Boltzmann constant and $T$ the temperature [32]. The 1DM results were computed at $T$=300 K. Eqs. (1) and (2) were numerically solved by means of a 4$^{th}$-order Runge-Kutta algorithm with a time step of 1ps over a temporal window of 100 ns.

The experimental results for the CIDWM are accurately reproduced by the 1DM predictions if we assume a DMI constant $D$=+0.06 mJ/m$^2$ and a SHA $\theta_{SH}$=-0.11, a pinning potential $V_0$=7x10$^{-20}$ J and $p$=21 nm as can be seen in Fig. 3(c), (d) and (e). The pinning potential parameters were selected to reproduce the experimentally observed propagation field (H$_p$≈5 Oe) in the absence of a current. It is interesting to note that the deduced positive DMI coefficient results in right-handed chiral Néel walls in the absence of a current and field (j$_a$=H$_x$=0). Therefore, the internal magnetization (**m**$_{DW}$) of an ↑↓- (↓↑-) domain wall points along the positive (negative) x-axis (**m**$_{DW}$≈±**u**$_x$). The extracted SHA value is in agreement with other values reported in literature [26,28].

Let us focus firstly on the ↑↓-domain wall case. Consistent with the negative value of the SHA for Ta, it is clear from Fig. 2 that the DW moves against the electron flow in the absence of longitudinal field. As already discussed, the effective field associated with the SHE is given by

$$\mathbf{H}_{SHE} = \frac{\hbar \theta_{SH}}{2e\mu_0 M_s L_z}[\mathbf{m}_{DW} \times (\mathbf{u}_z \times \mathbf{j}_a)] = \frac{\hbar \theta_{SH} j_a}{2e\mu_0 M_s L_z} m_{DW,x} \mathbf{u}_z = H_{SHE}\mathbf{u}_z,$$

which points along the easy axis (u$_z$) with a magnitude proportional to the x-component of the internal DW magnetization **m**$_{DW,x}$ (where $\mathbf{m}_{DW} = m_{DW,x}\mathbf{u}_x + m_{DW,y}\mathbf{u}_y$ is the internal DW magnetization in general). Fig. 3(d) indicates that for a given value of the applied current j$_a$, the ↑↓-DW remains pinned between an asymmetric longitudinal field range [H$_{d,-}$,H$_{d,+}$] with |H$_{d,-}$| > |H$_{d,+}$|: from μ$_0$H$_{d,-}$=-15 mT to μ$_0$H$_{d,+}$=-5 mT for j$_a$=3.6x10$^{12}$ A/m$^2$; and from μ$_0$H$_{d,-}$=-15 mT to μ$_0$H$_{d,+}$=0 mT for j$_a$=2.8x10$^{12}$ A/m$^2$. As a positive longitudinal field (H$_x$>0) is parallel to the internal DW magnetization of an ↑↓-DW at rest (m$_{DW,x}$>0), such a field stabilizes the Néel configuration of the DW against the initial rotation of **m**$_{DW}$ due to the current-induced torque, increasing the effective SHE field (H$_{SHE}$) and therefore supporting DW depinning. This results in a DW motion against the electron flow as in the absence of longitudinal field (e.g., with positive velocity for j$_a$>0). On the contrary, a negative longitudinal field ($H_x<0$) is anti-parallel to the internal DW magnetization of the right-handed ↑↓-DW at rest ($m_{DW,x}>0$), and consequently it reduces the magnitude of $m_{DW,x}$ and $H_{SHE}$. Therefore, $H_x<0$ acts against DW depinning which explains the experimental observation that for a given current $|H_{d,-}| > |H_{d,+}|$. The depinning field $\mu_\circ H_{d,-}$ is larger than the DMI effective field ($|\mu_\circ H_{d,-}| > \mu_\circ H_{DMI}=D/(M_S\Delta)$), so that when such a field is reached the direction of the DW motion reverses, being now along the electron flow. This explanation is also valid for the ↓↑-DW, where the internal DW magnetization is $m_{DW,x}<0$ at rest. In this case, the critical depinning values satisfy $|H_{d,+}| > |H_{d,-}|$ because it is now a negative longitudinal field ($H_x<0$) parallel to $m_{DW,x}$ which supports the DW depinning and subsequent propagation along the conventional current flow (see Fig. 3(c)).

## VII. CONCLUSIONS

Current-induced domain wall motion is observed in Ta\Co$_{20}$Fe$_{60}$B$_{20}$\MgO and Pt\Co$_{68}$Fe$_{22}$B$_{10}$\MgO nanowires, where the domain walls move against, respectively along the electron flow. The DW velocity is strongly influenced by a magnetic field applied along the wires. Moving from negative to positive applied magnetic fields the DW velocity is changed in its magnitude and in its direction, in agreement with the SOT-model. For the Ta sample a DMI-effective field $|\mu_o H_{DMI}|=+7.8\pm1.2$ mT is observed, resulting in a DMI constant $D=+0.06\pm0.01$ mJ/m$^2$. This corresponds to right-handed DWs in our materials system, in contrast to systems with CoFe. The positive DMI coefficient is attributed to the diffusion of boron in the tantalum buffer layer and its segregation at the Ta\Co$_{20}$Fe$_{60}$B$_{20}$ interface. Using 1D-model simulations we are able to reproduce the experimental data if we include pinning effects and we extract a spin-Hall angle of $\theta_{SH}=-0.11$ for Ta as well as the quantitative pinning strength.

**ACKNOWLEDGMENTS**

We acknowledge support by the Graduate School of Excellence Materials Science in Mainz (MAINZ) GSC 266, Staudinger Weg 9, 55128, Germany; the EU (IFOX, NMP3-LA-2012 246102;

MASPIC, ERC-2007-StG 208162; WALL, FP7-PEOPLE-2013-ITN 608031; MAGWIRE, FP7-ICT-2009-5), and the Research Center of Innovative and Emerging Materials at Johannes Gutenberg University (CINEMA). This work was also supported by EPSRC, U.K. (Grant Nos. EP/I011668/1, EP/K003127/1, EP/L00285X/1) and the Alexander von Humboldt Foundation CONNECT program. E. M. acknowledge the support by project MAT2011-28532-C03-01 from Spanish government and project SA163A12 from Junta de Castilla y Leon. Finally, we also thank the Singulus Technologies AG for support with the materials stacks preparation.

**FIGURE CAPTIONS**

FIG. 1. (a) Schematic of the experimental set-up for current pulse injection, including an SEM micrograph of the sample used during the experiment. The inset shows the shape of one of the pulses applied to the device, measured with the oscilloscope (across the 50 Ω internal resistance). (b) Differential Kerr microscopy image of nucleated magnetic domains by Oersted-field in initialized nanowires. The magnetization in the reversed domains is pointing upwards (+z, black areas). (c) Differential Kerr microscopy image of the same wires in (b), after domain walls motion by injecting a burst of 50 current pulses (Δt=25 ns, $j_a$=-2.75x10$^{11}$ A/m$^2$, $j_a$>0 in the +x direction) through the nanowires. The direction of the applied conventional current $j_a$ and of the electron current $j_e$ is indicated by the red and blue arrow respectively.

FIG. 2. (a) Average velocity of the DW as a function of the current density injected in the magnetic wires, for different durations of the current pulse. The DW velocity increases with the pulse duration, due to the fact that the 5 ns rise- and fall-time of the injected pulses have less influence on the measured domain wall velocity during longer pulses. The DW moves with the conventional current $j_a$ (against the electron flow $j_e$). The average velocities and the error bars (standard deviations) are calculated from 30 different DW motions, at each current density. (b) Average velocity of the DW as a function of $j_a$, free of the rise- and fall-time influence.

FIG. 3. Effect of a longitudinal magnetic field on the current-induced DW motion. (a) Differential Kerr microscopy image of nucleated magnetic domains in pre-saturated nanowires. The magnetization in the reversed domains points in the +z direction (black areas). The green lines indicate the position of the DWs. The red arrows describe the DWs magnetization configuration. (b) Differential Kerr microscopy image of the domain walls moved due to current pulse injection ($j_a$=+3.6x10$^{11}$ A/m$^2$), when a longitudinal field is applied ($\mu_\circ H_x$=-35 mT). The dashed green lines indicate the starting position of the DWs, while the solid orange lines indicate their final position. The blue arrows show the DW motion. Down-up (DU, ↓↑) and up-down (UD, ↑↓) DWs move in opposite direction. (c) Average velocity of ↓↑- and (d) ↑↓-DWs as a function of the longitudinal field ($\mu_\circ H_x$), for two different current densities. Solid symbols refer to $j_a$=3.6x10$^{11}$ A/m$^2$, while empty

symbols refer to $j_a=2.8\times10^{11}$ A/m$^2$. Squares refer to positive $j_a$, while triangles refer to negative $j_a$. The solid (dashed) lines are the 1D-model fitting-curves for $j_a=\pm3.6\times10^{11}$ A/m$^2$ ($j_a=\pm2.8\times10^{11}$ A/m$^2$) (see text for details). (e) Average velocity of ↓↑- (empty symbols) and ↑↓- (solid symbols) domain walls as a function of $\mu_\circ H_x$, for a current density of $j_a=+3.6\times10^{11}$ A/m$^2$ (squares), and $j_a=-3.6\times10^{11}$ A/m$^2$ (triangles). Lines represent the 1D-model fitting-curves.

FIG. 4. XMCD-PEEM images of current-induced DW motion in Pt\Co$_{68}$Fe$_{22}$B$_{10}$\MgO nanowires. (a) Image of the full device consisting in a magnetic nanowire with a magnetic pad at one side, connected to two gold pads at its ends (yellow areas) used for current pulses injection. The black line indicates the position of the DW in the nanowire. Part of the nanowire is not visible due to some residual resist from the patterning process. In (b), (c) and (d) the current-induced DW motion is shown. The dashed (solid) line indicates the initial (final) position of the DW. The DW moves each time with the electron flow, as indicated by the arrows.

FIG. 5. (a) TEM cross-section image of the Ta(5)\Co$_{20}$Fe$_{60}$B$_{20}$(1)\MgO(2)\Ta(5) stack showing that MgO and CoFeB crystallize in the cubic phase after annealing. Marks evidencing the different layers are superimposed as a guide to the eye. Layer thicknesses are reported in nm. (b) SIMS depth profiles of as-deposited (ad) and annealed (ann) (300°C, 2h) structures. Signals related to B (dots), MgO (squares), Fe (up-triangles) and Co (down-triangles) are shown. Following the B profile, the presence of B diffusion from Co$_{20}$Fe$_{60}$B$_{20}$ layer towards the Ta layer (and partially the MgO layer) is evidenced. For the sake of clarity profiles are aligned at Co$_{20}$Fe$_{60}$B$_{20}$\Ta interface. Secondary ions are collected in negative mode, and the measurement parameters are as reported in [36].

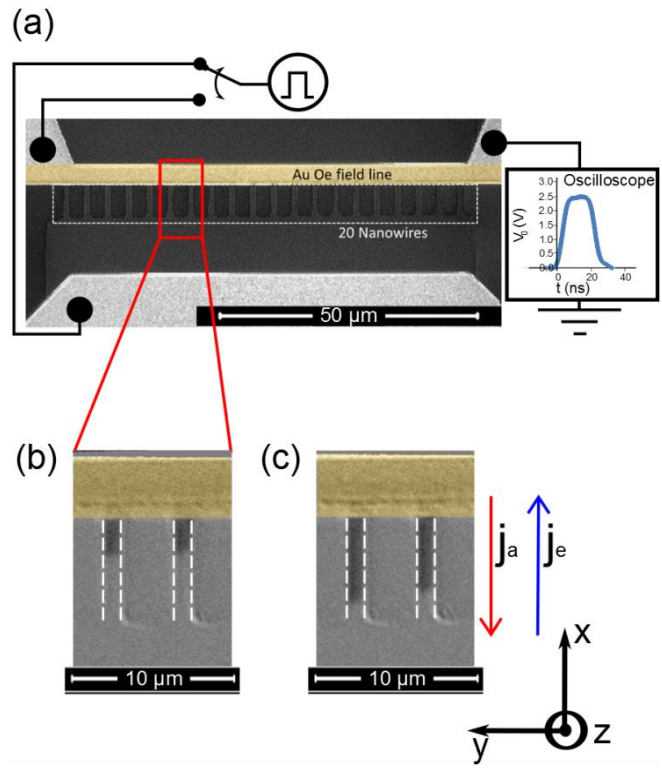

**Figure 1**

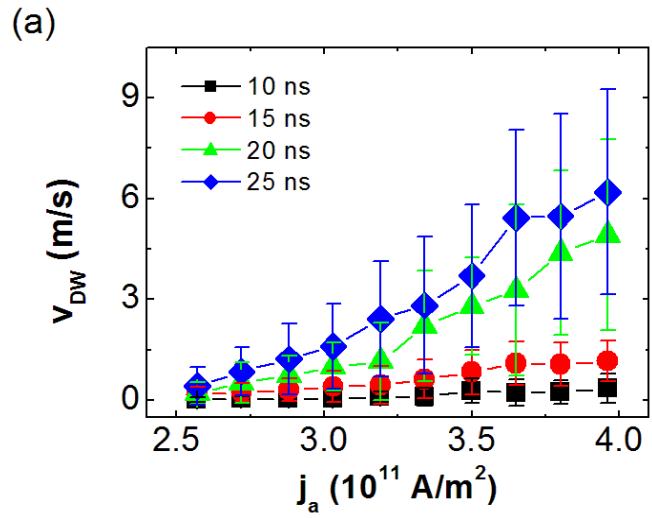

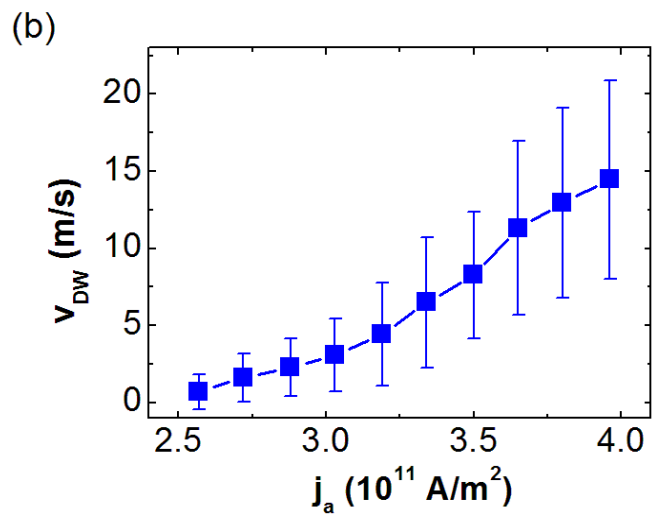

**Figure 2**

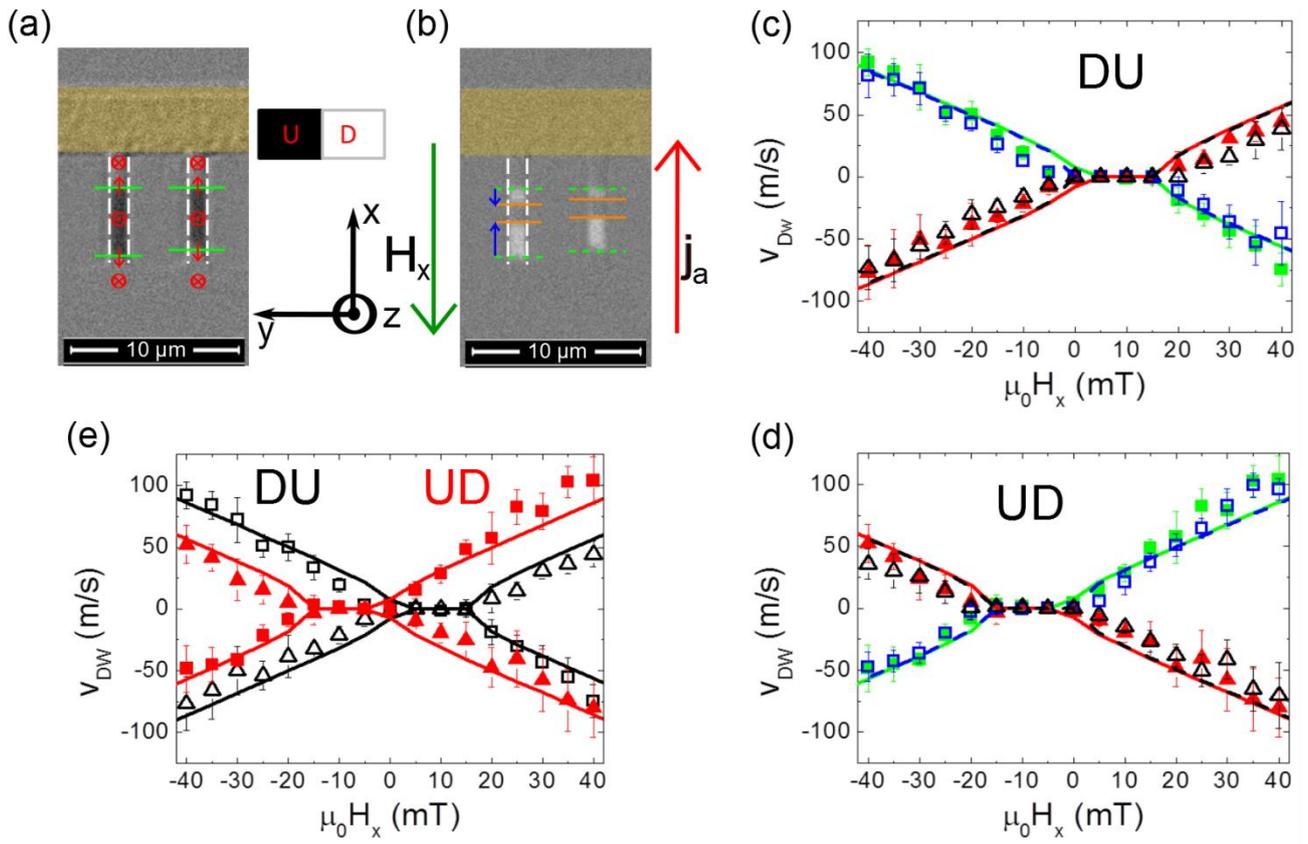

**Figure 3**

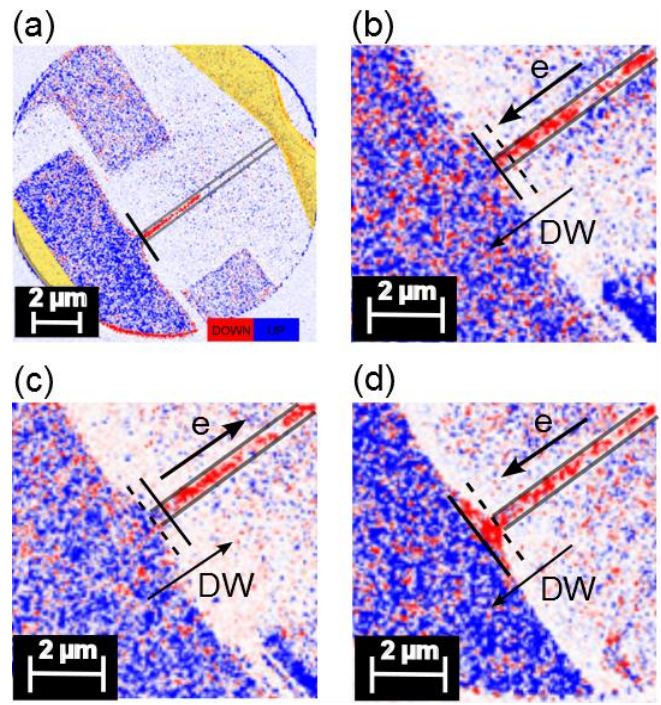

**Figure 4**

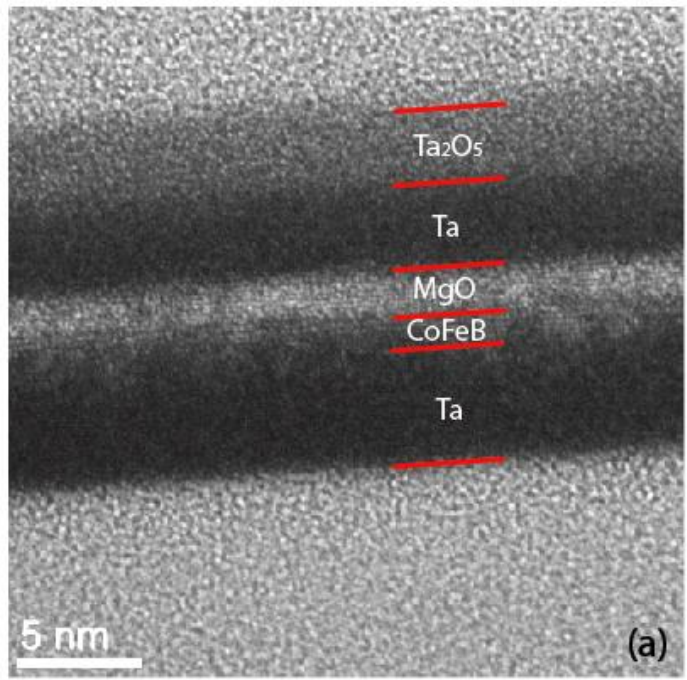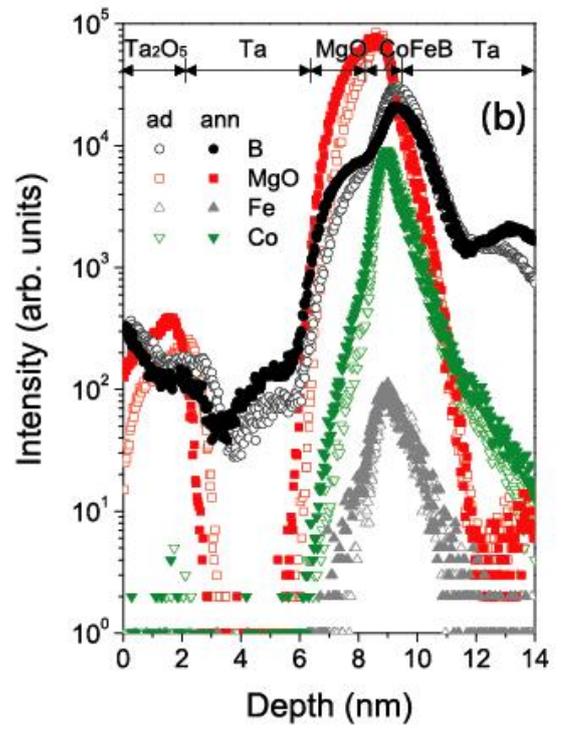

**Figure 5**